\documentstyle[11pt,newpasp,twoside,epsf]{article}
\def\edcomment#1{\iffalse\marginpar{\raggedright\sl#1\/}\else\relax\fi}
\reversemarginpar

\begin{document}
\title{Seismic tests of accretion in central stars of planetary systems}
 \author{Micha\"el Bazot}
\affil{Laboratoire d'Astrophysique, Observatoire Midi-Pyr\'en\'ees, 14 avenue Edouard Belin, 31400 Toulouse, France}
\author{Sylvie Vauclair}
\affil{Laboratoire d'Astrophysique, Observatoire Midi-Pyr\'en\'ees, 14 avenue Edouard Belin, 31400 Toulouse, France}

\begin{abstract}
Central stars of extra-solar planetary systems are metal-rich. Planet accretion or initial surmetallicity can explain this observationnal fact. These scenarios can be tested with asteroseismology. We calibrate two stellar models, one with accretion and one with high initial metallicity, in order to obtain the same external parameters for both of them. We then compare their internal structures and their oscillation frequencies.
\end{abstract}

\section{Introduction}

 Central stars of planetary systems are known to be overabundant in metals (see for example Santos et al. 2001; Santos et al. 2003). Two scenarios have been suggested to explain such an excess. In the first case, the protostellar gas is metal-rich and so is the star. The high density of heavy elements, compared to stars without planets, increase the probability of planetesimal formation, and thus of planet formation. The second explanation assumes that planets are formed out of gas with solar like metallicity. The central stars accrete hydrogene poor-matter during and after planetary formation thereby enhancing their metallicitiesin their outer layer (Murray et al. 2001). We suggest that these two scenarios can be tested by asteroseismology. Here we analyse and compare the oscillation frequencies in stellar models with the same external parameters but different histories : with and without accretion.
\section{Calibration}
We computed 1.1 M$_{\odot}$ stars models with the Toulouse-Geneva code of stellar evolution. One of them has a high initial metallicity, from the surface to the core, Z=0.0443. The other model has initial solar abundances but it experienced accretion at the beginning of its main-sequence evolution. Both models include microscopic diffusion. We adjusted the mass of accreted materia (M$_a$), the mixing-length parameter ($\alpha$) and the initial helium abundance (Y$_0$), which are input parameters in our stellar evolution code, in order to obtain stars with the same effective temperature (T$_{eff}$), luminosity (L) and metallicity (Z). Input and output parameters for these models are given in table 1 and table 2, evolution tracks are represented in figure 1.\\

\begin{figure}
\plotone{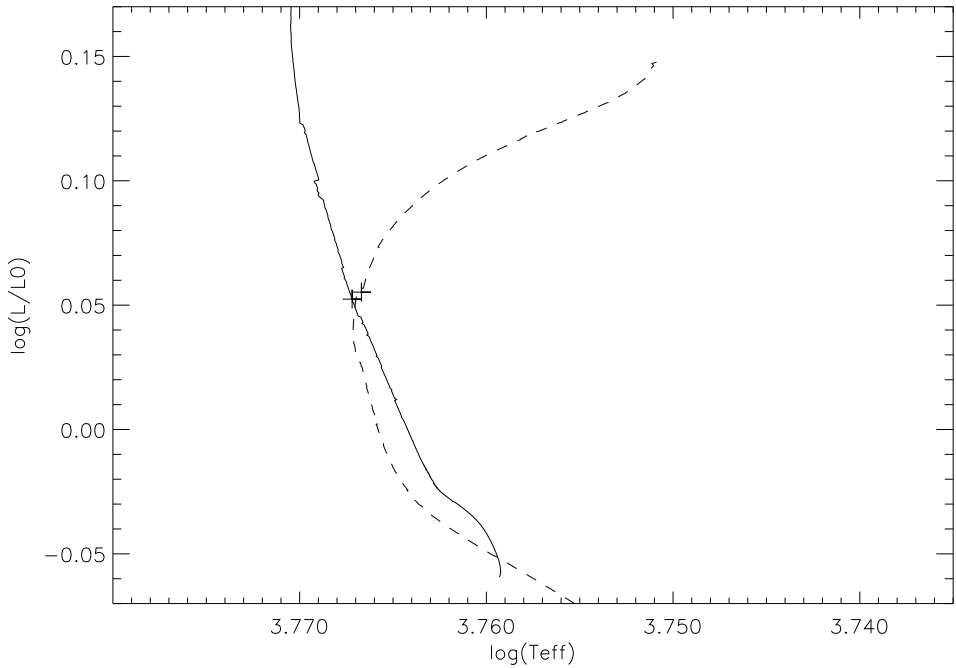}
\caption{Evolutionnary tracks for a star experiencing accretion of planetary material (full line) and supermetallic star (dashed line) The models used for this study are marked with a cross on each track.}
\end{figure} 

 The accreted mass is composed of elements heavier than helium with the same relatives abundances as in the Sun. It is given in units of M$_{Jup}$ (M$_{Jup}$ = 1.9$\times$10$^{27}$ kg). When the convergence of two models is obtained, we compute their oscillation frequencies. We used the adiabatic pulsations code described by Brassard et al. (1992).

\begin{table}[h]

\begin{tabular}{|l|c|c|c|c|c|c|}
\tableline
Model & M$_{\star}$ (M$_{\odot}$) &  M$_a$ (M$_{Jup}$) & $\alpha$ & X$_0$ & Y$_0$ &  Z$_0$  \\
\tableline
1 & 1.1 & 0.5 & 2.08 & 0.7208 & 0.2600 & 0.0192  \\
2 & 1.1 & - & 2.08 & 0.6260 & 0.3297 & 0.0443 \\
\tableline
\tableline  

\end{tabular}
\caption{Intput parameters for model 1 (with accretion) and model 2 (supermetallic).The mass fractions X$_0$, Y$_0$, Z$_0$ are the surface quantities. $\alpha$ is the mixing-length parameter. }
\end{table}

\begin{table}[h]

\begin{tabular}{|l|c|c|c|c|c|c|c|c|c|}
\tableline
Model  & Age (Gyr) & X & Y & Z & T$_{eff}$ (K) & $\frac{L}{L_{\odot}}$ & $\frac{R}{R_{\odot}}$ & $\frac{R_{cv}}{R_{\star}}$ & C.C. \\
\tableline
1 & 2.123 & 0.7213 & 0.2389 & 0.0482 & 5850 & 1.128 & 1.0368 & 0.7085 & no \\
2 &  2.965 & 0.6541 & 0.3056 & 0.0403 & 5847 & 1.136 & 1.0415 & 0.7035 & yes\\
\tableline
\tableline  

\end{tabular}
\caption{Output parameters for model 1 (with accretion) and model 2 (supermetallic). The mass fractions X, Y, Z are the surface quantities. The last column indicates whether or not the star has a convective core.}
\end{table}

\section{Results}
\subsection{Physical parameters of the stars}
 As the physical processes in the stars are not the same, the internal structures of models with the same external parameters present detectable differences. In figure 2a we plotted the relative differences of temperature, pression, density and the square of the sound speed for the point on the HR diagram (see figure 1)  at which the stars have similar external parameters. In table 2 $R_{cv}$ is the radius at the bottom of the outer convective zone, R$_{\star}$ is the star radius. The last column indicates whether or not the star has a convective core. 

\begin{figure}[h]
\plottwo{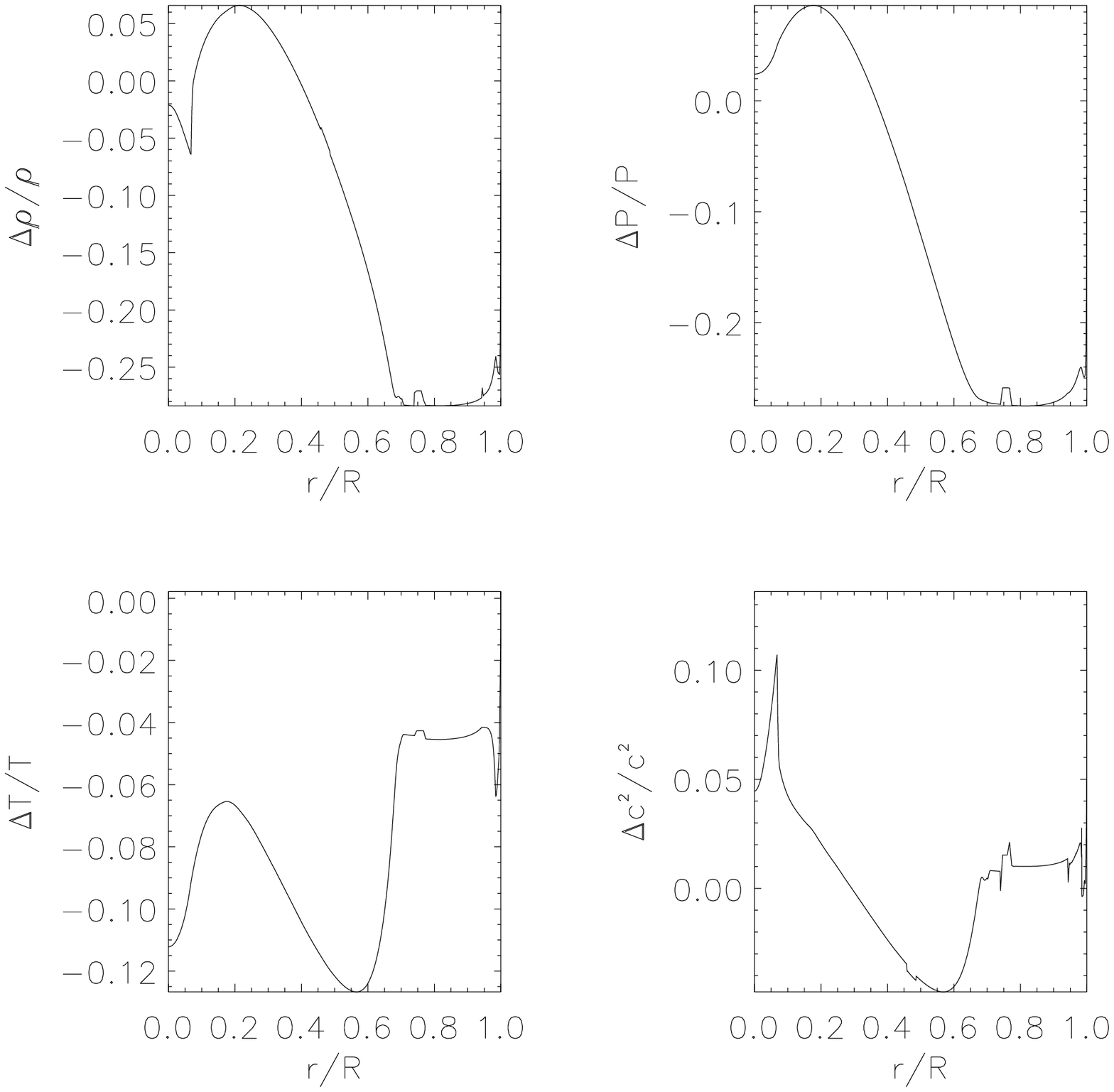}{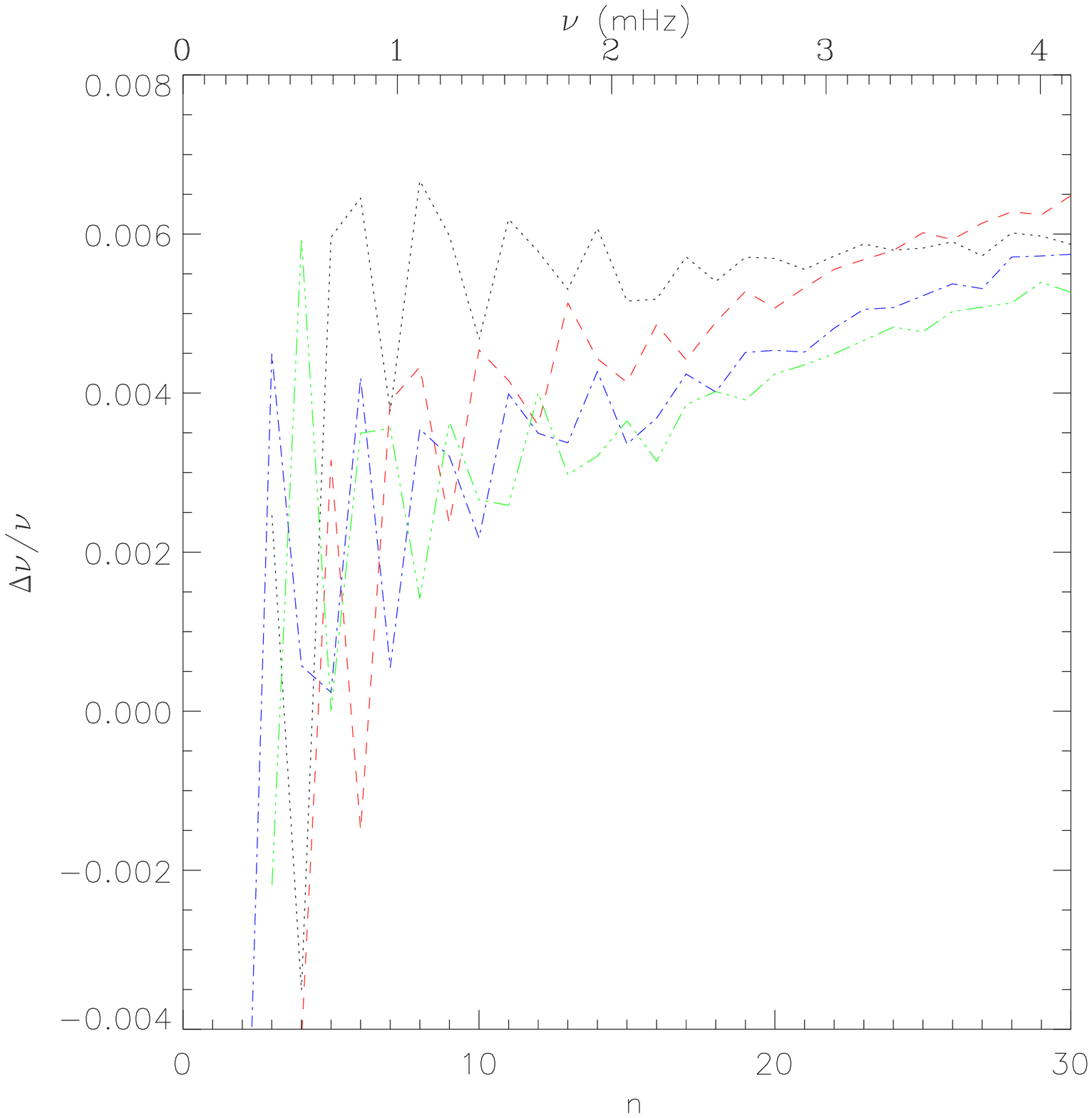}
\caption{ {\bf a}. Relative differences in density, pression, temperature and sound speed. For example $\frac{\Delta T}{T}$ stand for $(T_{accr}-T_{sm})/T_{sm})$ where the subscript {\it accr} describes the star with accretion and {\it sm} the supermetallic one. {\bf b}.  Relative differences of the oscillation frequencies for azimuthal degrees l=0 (full line), 1 (dot-dashed line), 2 (dashed line) and 3 (double dotted-dashed line).}
\end{figure} 

\subsection{Oscillation frequencies}
We computed the oscillation frequencies of our models for different values of the azimuthal degree {\it l} (from 0 to 3) and radial order {\it n}. We then compared the oscillation frequencies. The relative difference between oscillation frequencies, as functions of the radial order and the frequency itself, is shown in figure 3.

\section{Discussion}
 Sound speed differences, which appear in figure 2a, are worth studying. It reaches 0.5 \% below the convective zone and 1 \% near the center. In the first case, it reflects the difference in temperature and chemical abundances. The mean molecular weight, $\mu$, differs and so does the sound speed. The peak at $\sim$ 0.05 R$_{\star}$ is due to the presence of a convective core in the supermetallic star. Figure 2b pictures the differences of adiabatic oscillation frequencies linked to the differences in sound speed. They are of the order of a few tenth of percent which is an observable effect with the means of asteroseismology (Bouchy \& Carrier 2002).\\
 A further level in analysis consists in computing the small and second differences for both stars, i.e. $\delta \nu_{n,l}=\nu_{nl}-\nu_{n-1,l+2}$ and $\delta_{2} \nu=\nu_{n+1}-2\nu_{n}+\nu_{n-1}$. The former is a good test for stellar interior near the core (Ulrich 1986). The latter is a good indicator for the region near the convective zone (Gough 1990). Computation of these quantities are under process, they will discussed in a forthcoming paper.\\
\\

\end{document}